\pdfoutput=1
\documentclass[prl,letterpaper,twocolumn,10pt,aps,notitlepage]{revtex4-1}

\usepackage{amsfonts}
\usepackage{amsmath}
\usepackage{bm,graphicx}
\usepackage{hyperref}
\usepackage{placeins}

\hypersetup{colorlinks, linkcolor = [rgb]{0,0.0,0.75}, citecolor = [rgb]{0,0.0,0.75}, urlcolor = [rgb]{0,0.0,0.75}}

\usepackage[utf8]{inputenc}

\newcommand{\nb}{\phantom{0}}
\newcommand{\wm}{\phantom{-}}

\begin{document}

\title{\texorpdfstring{$\bm{\Lambda_c\to\Lambda \ell^+\nu_\ell}$}{Lambda_c to Lambda l^+ nu_l}
form factors and decay rates from lattice QCD with physical quark masses}

\author{Stefan Meinel}
\affiliation{Department of Physics, University of Arizona, Tucson, AZ 85721, USA}
\affiliation{RIKEN BNL Research Center, Brookhaven National Laboratory, Upton, NY 11973, USA}

\date{February 21, 2017}

\begin{abstract}
The first lattice QCD calculation of the form factors governing $\Lambda_c \to \Lambda \ell^+ \nu_\ell$ decays is reported. The calculation was performed with two different lattice spacings and includes one ensemble with a pion mass of 139(2) MeV. The resulting predictions for the $\Lambda_c \to \Lambda e^+ \nu_e$ and $\Lambda_c \to \Lambda \mu^+ \nu_\mu$ decay rates divided by $|V_{cs}|^2$ are $0.2007(71)(74)\:{\rm ps}^{-1}$ and $0.1945(69)(72)\:{\rm ps}^{-1}$, respectively, where the two uncertainties are statistical and systematic. Taking the Cabibbo-Kobayashi-Maskawa matrix element $|V_{cs}|$ from a global fit and the $\Lambda_c$ lifetime from experiments, this translates to branching fractions of $\mathcal{B}(\Lambda_c\to\Lambda e^+\nu_e)=0.0380(19)_{\rm LQCD}(11)_{\tau_{\Lambda_c}}$ and $\mathcal{B}(\Lambda_c\to\Lambda \mu^+\nu_\mu)=0.0369(19)_{\rm LQCD}(11)_{\tau_{\Lambda_c}}$. These results are consistent with, and two times more precise than, the measurements performed recently by the BESIII Collaboration. Using instead the measured branching fractions together with the lattice calculation to determine the CKM matrix element gives $|V_{cs}|= 0.949(24)_{\rm LQCD}(14)_{\tau_{\Lambda_c}}(49)_{\mathcal{B}}$.
\end{abstract}

\maketitle

Precision studies of processes in which heavy bottom or charm quarks decay to lighter quarks play an important role
in testing the Standard Model of elementary particle physics. While most of these analyses are being performed using $B$ and $D$ mesons,
decays of $\Lambda_b$ and $\Lambda_c$ baryons can provide valuable additional information. Two examples that shed new
light on puzzles posed by mesonic decays are the determination of the ratio of CKM matrix elements $|V_{ub}/V_{cb}|$
from the $\Lambda_b \to p \mu^-\bar{\nu}_\mu$ and $\Lambda_b \to \Lambda_c \mu^-\bar{\nu}_\mu$ decay rates \cite{Aaij:2015bfa},
and an analysis of the rare $b\to s \mu^+\mu^-$ transition using $\Lambda_b \to \Lambda \mu^+\mu^-$ \cite{Meinel:2016grj}. Both
studies rely on nonperturbative lattice QCD calculations of form factors describing the baryonic matrix elements of the underlying
quark currents \cite{Detmold:2015aaa, Detmold:2016pkz}.

This letter focuses on the charmed-baryon decays $\Lambda_c \to \Lambda \ell^+ \nu_\ell$ ($\ell=e,\mu$),
whose rates are proportional to $|V_{cs}|^2$ in the Standard Model. Previous determinations of this CKM matrix element are
\begin{equation}
 \hspace{-1ex}|V_{cs}| = \left\{\begin{array}{ll} 1.008(5)(16) & \text{from } D_s \to \ell^+\nu_\ell \text{ \cite{Bazavov:2014wgs, Aoki:2016frl}}, \\
                                     0.975(25)(7) & \text{from } D \to K \ell^+\nu_\ell \text{ \cite{Na:2010uf, Aoki:2016frl}}, \\
                                     0.97344(15)  & \text{indirect, CKM unitarity \cite{UTfit}.} \end{array}\right. \hspace{-1ex} \label{eq:Vcsprevious}
\end{equation}
The motivations for studying $\Lambda_c \to \Lambda \ell^+ \nu_\ell$ include the following:
\begin{enumerate}
 \item Taking the precisely determined value of $|V_{cs}|$ from CKM unitarity, a comparison between calculated and measured
       $\Lambda_c \to \Lambda \ell^+ \nu_\ell$ decay rates provides a stringent test of the methods used to compute the heavy-baryon decay form factors.
 \item Combining the $\Lambda_c \to \Lambda \ell^+ \nu_\ell$ decay rates from experiment with a lattice QCD calculation of the $\Lambda_c \to \Lambda$
       form factors gives a new direct determination of $|V_{cs}|$ and new constraints on physics beyond the Standard Model
       (see, e.g., Ref.~\cite{Fajfer:2015ixa} for a recent discussion of new physics in $c \to s \ell^+ \nu_\ell$ transitions).
 \item If the $\Lambda_c \to \Lambda \ell^+ \nu_\ell$ decay rates are known precisely, from experiment or lattice QCD,
       these modes can be used as normalization modes in measurements of a wide range of other charm and bottom baryon decays \cite{Rosner:2012gj}.
\end{enumerate}
The most precise measurements of the $\Lambda_c \to \Lambda \ell^+ \nu_\ell$ branching fractions (decay rates times the $\Lambda_c$ lifetime)
to date have recently been reported by the BESIII Collaboration \cite{Ablikim:2015prg, Ablikim:2016vqd},
\begin{equation}
 \mathcal{B}(\Lambda_c\to\Lambda \ell^+\nu_\ell) =
 \left\{\begin{array}{ll} 0.0363(38)(20), &\ell=e, \\ 
                          0.0349(46)(27), & \ell=\mu. \end{array} \right. \label{eq:BESIII}
\end{equation}
In the Standard Model, the decay rates depend on six form factors that parametrize
the matrix elements $\langle \Lambda(p^\prime) |\: \bar{s}\gamma^\mu c\: |\Lambda_c(p)\rangle$
and $\langle \Lambda(p^\prime) | \: \bar{s}\gamma^\mu\gamma_5 c\: |\Lambda_c(p)\rangle$
as functions of $q^2=(p-p^\prime)^2$. These form factors have previously been estimated using quark models and sum rules
\cite{Buras:1976dg, Gavela:1979wk, AvilaAoki:1989yi, PerezMarcial:1989yh, Hussain:1990ai, Singleton:1990ye, Efimov:1991ex,Garcia:1992qe,
Cheng:1995fe, Ivanov:1996fj, Dosch:1997zx, Luo:1998wg, MarquesdeCarvalho:1999bqs, Pervin:2005ve, Liu:2009sn, Gutsche:2015rrt, Faustov:2016yza},
giving branching fractions that vary substantially depending on the model assumptions. In the following, the first lattice QCD determination of the $\Lambda_c \to \Lambda$
form factors is reported. The calculation uses state-of-the-art methods and gives predictions for the $\Lambda_c \to \Lambda \ell^+ \nu_\ell$ decay rates
with total uncertainties that are smaller than the experimental uncertainties in Eq.~(\ref{eq:BESIII}) by a factor of two.

\begin{table*}
\begin{tabular}{ccccccccccccccccccccc}
\hline\hline
Set & \hspace{1ex} & $\beta$ & \hspace{1ex} & $N_s^3\times N_t$ & \hspace{1ex} & $am_{u,d}^{(\mathrm{sea})}$
& \hspace{1ex} & $am_{s}^{(\mathrm{sea})}$   & \hspace{1ex} & $a$ [fm] & \hspace{1ex} & $am_{u,d}^{(\mathrm{val})}$ 
& \hspace{1ex} & $m_\pi^{(\mathrm{val})}$ [MeV] & \hspace{1ex} & $am_{s}^{(\mathrm{val})}$ 
& \hspace{1ex} & $m_{\eta_s}^{(\mathrm{val})}$ [MeV] & \hspace{1ex} & $N_{\rm samples}$ \\
\hline
\texttt{CP}  && $2.13$ && $48^3\times96$ && $0.00078$ && $0.0362$ && $0.1142(15)$  && $0.00078$ && 139(2) && $0.0362$ && 693(9)  && 2560 sl, 80 ex \\
\texttt{C54} && $2.13$ && $24^3\times64$ && $0.005$   && $0.04$   && $0.1119(17)$  && $0.005$   && 336(5) && $0.04$   && 761(12) && 2782 \\
\texttt{C53} && $2.13$ && $24^3\times64$ && $0.005$   && $0.04$   && $0.1119(17)$  && $0.005$   && 336(5) && $0.03$   && 665(10) && 1205 \\
\texttt{F43} && $2.25$ && $32^3\times64$ && $0.004$   && $0.03$   && $0.0849(12)$  && $0.004$   && 295(4) && $0.03$   && 747(10) && 1917 \\
\texttt{F63} && $2.25$ && $32^3\times64$ && $0.006$   && $0.03$   && $0.0848(17)$  && $0.006$   && 352(7) && $0.03$   && 749(14) && 2782 \\
\hline\hline
\end{tabular}
\caption{\label{tab:params}Parameters of the lattice gauge field ensembles and $u$, $d$, $s$ quark propagators \cite{Blum:2014tka, Aoki:2010dy}. The
lattice spacings given here were determined using the $\Upsilon(2S)-\Upsilon(1S)$ splitting \cite{Meinel:2010pv}.
The $\eta_s$ is an artificial pseudoscalar $s\bar{s}$ meson used to tune the strange-quark mass \cite{Davies:2009tsa}; at the physical point,
one has $m_{\eta_s}^{(\mathrm{phys})}=689.3(1.2)\:\:{\rm MeV}$ \cite{Dowdall:2011wh}. On the \texttt{CP} ensemble,
\emph{all-mode-averaging} \cite{Shintani:2014vja} with 64 sloppy (sl) and 2 exact (ex) samples per gauge configuration was used for the computation of the
quark propagators. }
\end{table*}

\begin{table}
\begin{tabular}{ccccccccc}
\hline\hline
Set  & \hspace{1ex} & $a m_{\Lambda_c}$   & \hspace{1ex} & $a m_{\Lambda}$  & \hspace{1ex} & $am_{D_s}$ & \hspace{1ex} & $am_{D}$    \\
\hline
\texttt{CP}         && $1.3194(36)$     &&  $0.6483(33)$     && $1.12902(39)$     &&  $1.0720(12)\nb$   \\
\texttt{C54}        && $1.3706(40)$     &&  $0.7348(30)$     && $1.13156(49)$     &&  $1.0763(13)\nb$   \\
\texttt{C53}        && $1.3647(60)$     &&  $0.7096(47)$     && $1.11550(59)$     &&  $1.0763(13)\nb$   \\
\texttt{F43}        && $1.0185(67)$     &&  $0.5354(29)$     && $0.85447(47)$     &&  $0.81185(91)$     \\
\texttt{F63}        && $1.0314(40)$     &&  $0.5514(23)$     && $0.85639(33)$     &&  $0.81722(56)$     \\
\hline\hline
\end{tabular}
\caption{\label{tab:hadronmasses}Hadron masses in lattice units obtained from exponential fits to two-point functions.}
\end{table}

This work is based on gauge field configurations generated by the RBC and UKQCD collaborations with $2+1$ flavors of dynamical
domain-wall fermions \cite{Aoki:2010dy, Blum:2014tka}. The data sets used here are listed in Table \ref{tab:params}, and match those in
Refs.~\cite{Detmold:2015aaa} and \cite{Detmold:2016pkz}, except for the addition of a new ensemble (denoted as \texttt{CP}) with $m_\pi=139(2)$ MeV,
and the removal of the previous ``partially quenched'' \texttt{C14}, \texttt{C24}, \texttt{F23} data sets which had
$am_{u,d}^{(\mathrm{val})}<am_{u,d}^{(\mathrm{sea})}$. Adding the \texttt{CP} ensemble significantly aids in the extrapolation of the form
factors to the physical point, and removing the partially quenched data sets reduces finite-volume effects.

The charm quark is implemented using an anisotropic clover action, with parameters tuned
to produce the correct $J/\psi$ relativistic dispersion relation as quantified by the ``speed of light'', $c$, and the correct
spin-averaged mass $\overline{m}=\frac34m_{J/\psi}+\frac14m_{\eta_c}$ \cite{Brown:2014ena}. On the new \texttt{CP} ensemble,
the same bare parameters as tuned on the coarse $24^3\times64$ lattice yield $c=0.9970(27)$ and $\overline{m}=3019(40)$ MeV,
consistent with the experimental value of $3068.5(0.1)$ MeV \cite{Olive:2016xmw}, and were therefore used on this ensemble as well.
The $\Lambda_c$, $\Lambda$, $D_s$, and $D$ masses obtained from the different data sets are listed in Table \ref{tab:hadronmasses}.

The renormalization of the $c\to s$ vector and axial vector currents is performed using the \emph{mostly nonperturbative} method
\cite{Hashimoto:1999yp, ElKhadra:2001rv} as in Eqs.~(18)-(21) of Ref.~\cite{Detmold:2015aaa} (with the replacements $b\to c$, $q\to s$).
The nonperturbative coefficients used here on the coarse $48^3\times96$, coarse $24^3\times64$, and fine $32^3\times64$ lattices are
$Z_V^{(ss)}=0.71076(25),0.71273(26),0.74404(181)$ \cite{Blum:2014tka} and $Z_V^{(cc)}=1.35899(22),1.35725(23),1.18321(14)$;
the residual matching coefficients and $\mathcal{O}(a)$-improvement coefficients were computed in tadpole-improved one-loop
lattice perturbation theory \cite{Lehner:2012bt, Lehnercharmlight} and are given in Table \ref{tab:Pmatchingfactors}.

\begin{table}
\begin{center}
\small
\begin{tabular}{cllll}
\hline\hline
Parameter          & \hspace{2ex} & \hspace{1ex} Coarse lattice    & \hspace{2ex} &  \hspace{1ex} Fine lattice       \\
\hline
 $\rho_{V^0}=\rho_{A^0}$     && $\wm1.00274(49)$      &&  $\wm1.001949(85)$   \\[0.2ex]
 $\rho_{V^j}=\rho_{A^j}$     && $\wm0.99475(62)$      &&  $\wm0.99675(68)$    \\[0.2ex]
 $c_{V^0}^R=c_{A^0}^R$       && $\wm0.0402(88)$       &&  $\wm0.0353(92)$     \\[0.2ex]
 $c_{V^0}^L=c_{A^0}^L$       && $-0.0048(48)$         &&  $-0.0027(28)$       \\[0.2ex]
 $c_{V^j}^R=c_{A^j}^R$       && $\wm0.0346(51)$       &&  $\wm0.0283(43)$     \\[0.2ex]
 $c_{V^j}^L=c_{A^j}^L$       && $\wm0.00012(26)$      &&  $\wm0.00040(42)$    \\[0.2ex]
 $d_{V^j}^R=-d_{A^j}^R$      && $-0.0041(41)$         &&  $-0.0039(39)$       \\[0.2ex]
 $d_{V^j}^L=-d_{A^j}^L$      && $\wm0.0021(21)$       &&  $\wm0.0026(26)$     \\[0.2ex]
\hline\hline
\end{tabular}\vspace{-2ex}
\end{center}
\caption{\label{tab:Pmatchingfactors} Residual matching and improvement coefficients for the $c\to s$ vector and axial vector currents,
computed using automated lattice perturbation theory \cite{Lehner:2012bt, Lehnercharmlight}.
The notation is the same as in Eqs.~(18)-(21) of Ref.~\cite{Detmold:2015aaa}.}
\end{table}

\begin{figure*}
\begin{center}
 \hfill \includegraphics[width=0.46\linewidth]{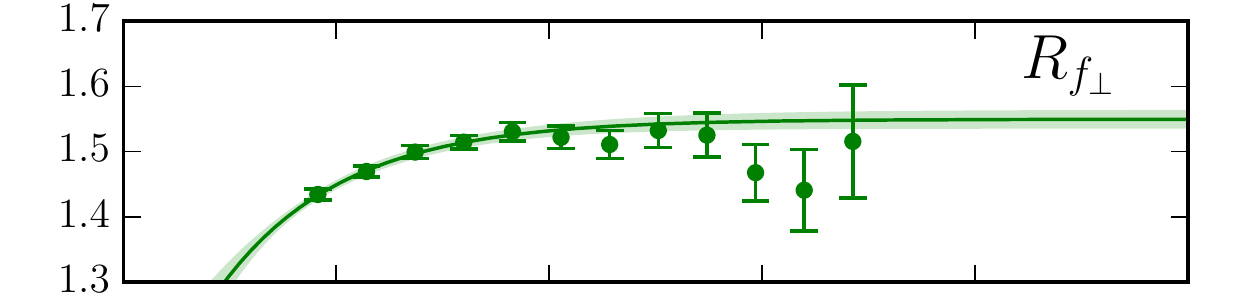} \hfill \includegraphics[width=0.46\linewidth]{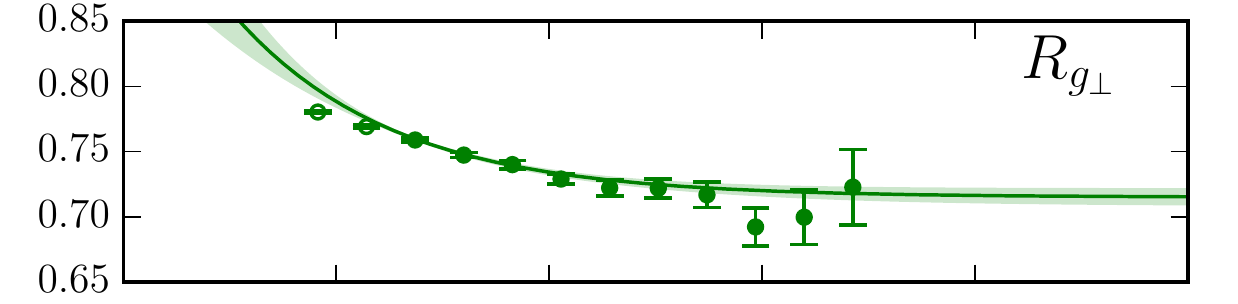} \hfill \null \\
 \hfill \includegraphics[width=0.46\linewidth]{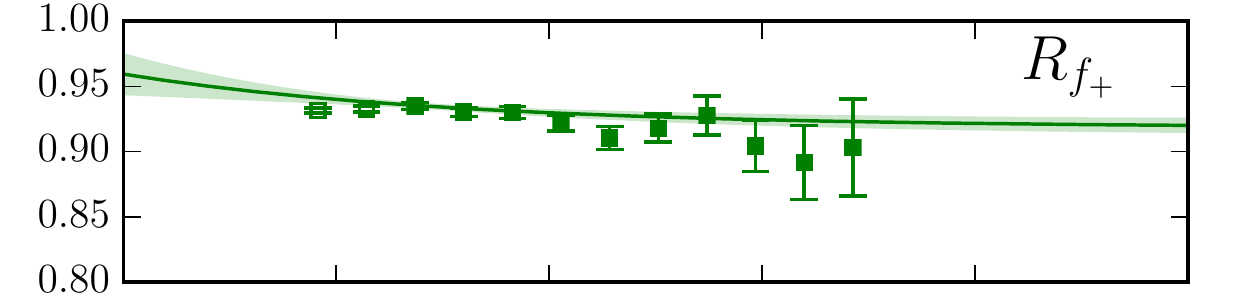} \hfill \includegraphics[width=0.46\linewidth]{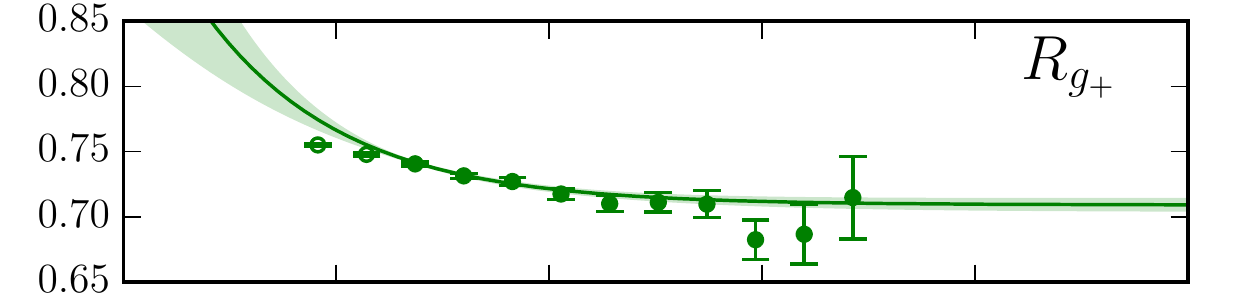} \hfill \null \\
 \hfill \includegraphics[width=0.46\linewidth]{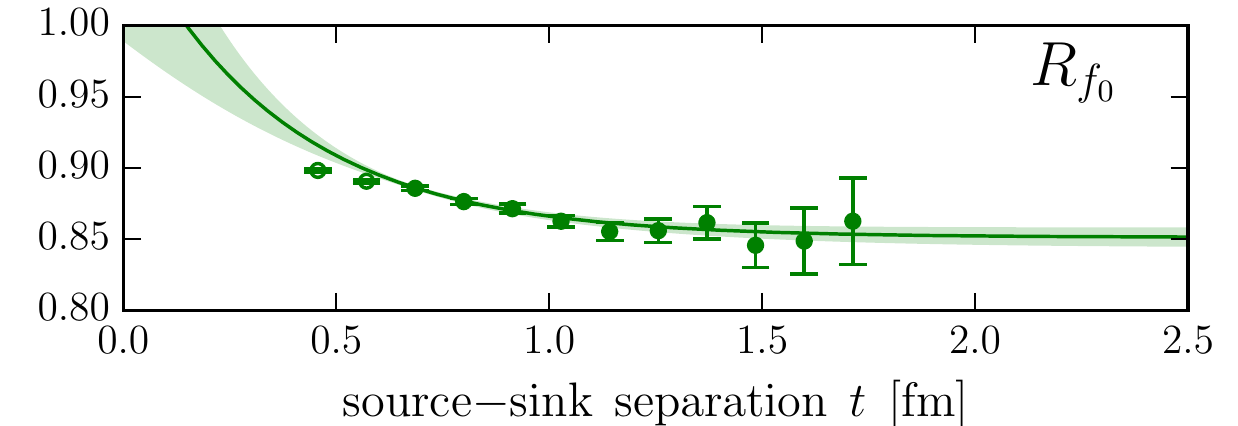} \hfill \includegraphics[width=0.46\linewidth]{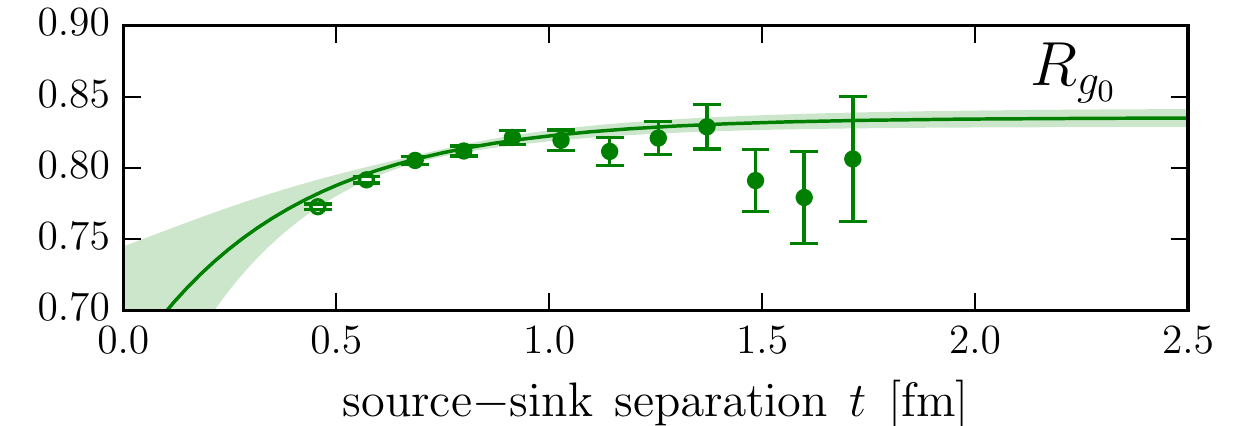} \hfill \null \\
 \caption{\label{fig:tsepextrap}Lattice results from the $\texttt{CP}$ ensemble for the ratios $R_f(|\mathbf{p}^\prime|,t)$,
 defined as in Eqs.~(52-54) and (58-60) of Ref.~\cite{Detmold:2015aaa}, at $|\mathbf{p}^\prime|^2=4(2\pi/L)^2$. These ratios
 are equal to the ground-state form factors $f(|\mathbf{p}^\prime|)$ up to contamination from excited states that decays exponentially with $t$.
 The curves shown are correlated fits of the form $R_f(|\mathbf{p}^\prime|,t)=f(|\mathbf{p}^\prime|)+A_f(|\mathbf{p}^\prime|)\, e^{-\delta_f(|\mathbf{p}^\prime|)\,t}$, which
 includes the leading excited-state contributions. Data points at the smallest separations that are plotted with open symbols are excluded from the
 fits to suppress contamination from higher excited states; the values of $t_{\rm min}$ were chosen such that $\chi^2/{\rm d.o.f.}\leq 1$. The remaining systematic uncertainties due to higher excited states
 were estimated as the shifts in the fitted $f(|\mathbf{p}^\prime|)$ when further increasing $t_{\rm min}$ by one unit everywhere; these uncertainties were added in quadrature to the
 statistical uncertainties.} 
\end{center}
\end{figure*}

The $\Lambda_c\to\Lambda$ form factors are defined as in Eqs.~(1) and (2) of Ref.~\cite{Detmold:2016pkz} (with $b$ replaced by $c$), and were
extracted from ratios of three-point and two-point correlation functions using the same methods as in \cite{Detmold:2015aaa,Detmold:2016pkz}. This
involves extrapolations to infinite source-sink separation to isolate the ground-state contributions, which are performed jointly for
all data sets at matching $\Lambda$ momenta \cite{Detmold:2015aaa,Detmold:2016pkz}. These momenta, $|\mathbf{p}^\prime|^2$,
were set to 1, 2, 3, and 4 times $(2\pi/L)^2$ on all data sets except \texttt{CP}. For the latter, the values 4, 8, 12, and 16 times $(2\pi/L)^2$
were used because $L=N_s a$ is twice as large. The ranges of source-sink separations were $t/a=4...15$ on the coarse lattices and $t/a=5...17$ on the fine lattices;
full $\mathcal{O}(a)$-improvement of the currents was performed for all source-sink separations (instead of just a subset as in
Refs.~\cite{Detmold:2015aaa,Detmold:2016pkz}). Examples for the ratios and $t\to\infty$ extrapolations are shown in Fig.~\ref{fig:tsepextrap}.

\begin{figure*}
\includegraphics[width=0.9\linewidth]{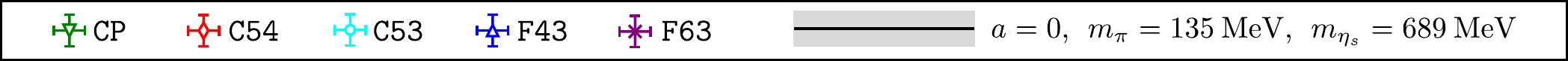}
\vspace{2ex}
\includegraphics[width=0.45\linewidth]{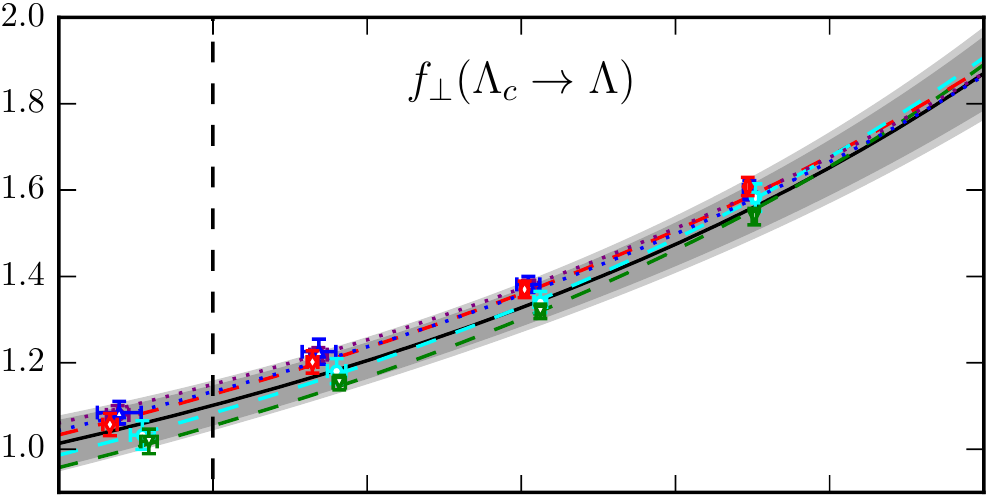} \includegraphics[width=0.45\linewidth]{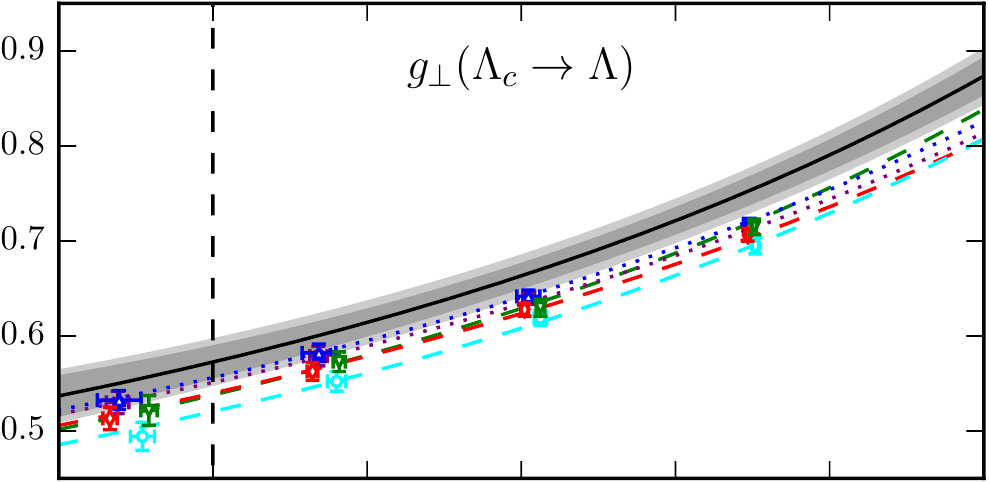} 
\includegraphics[width=0.45\linewidth]{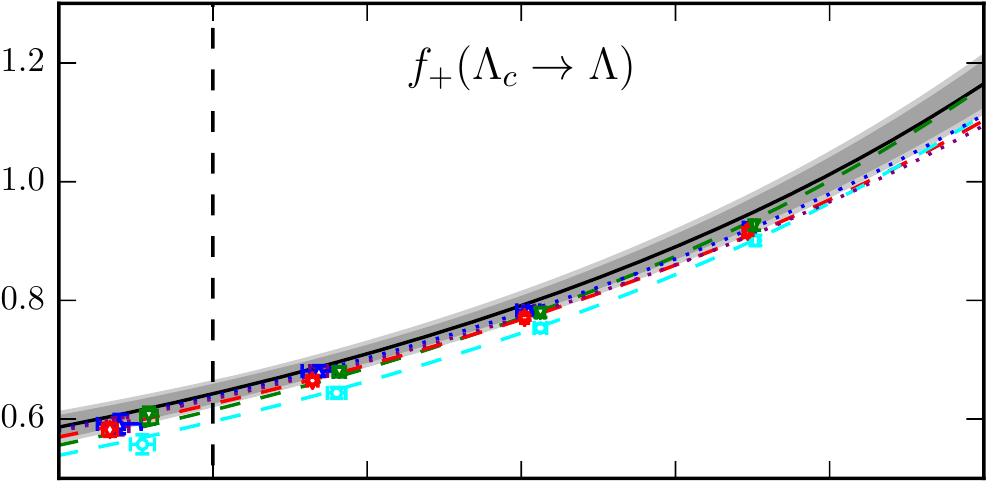} \includegraphics[width=0.45\linewidth]{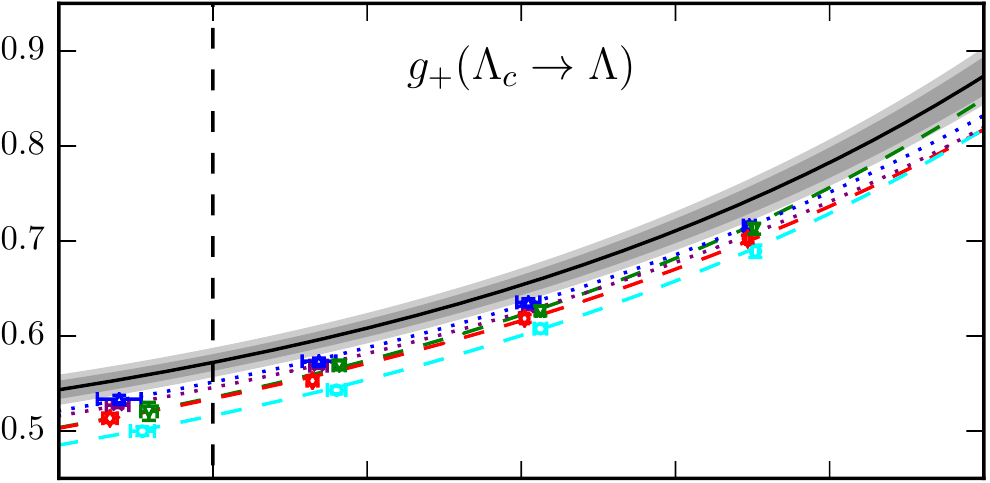} 
\includegraphics[width=0.45\linewidth]{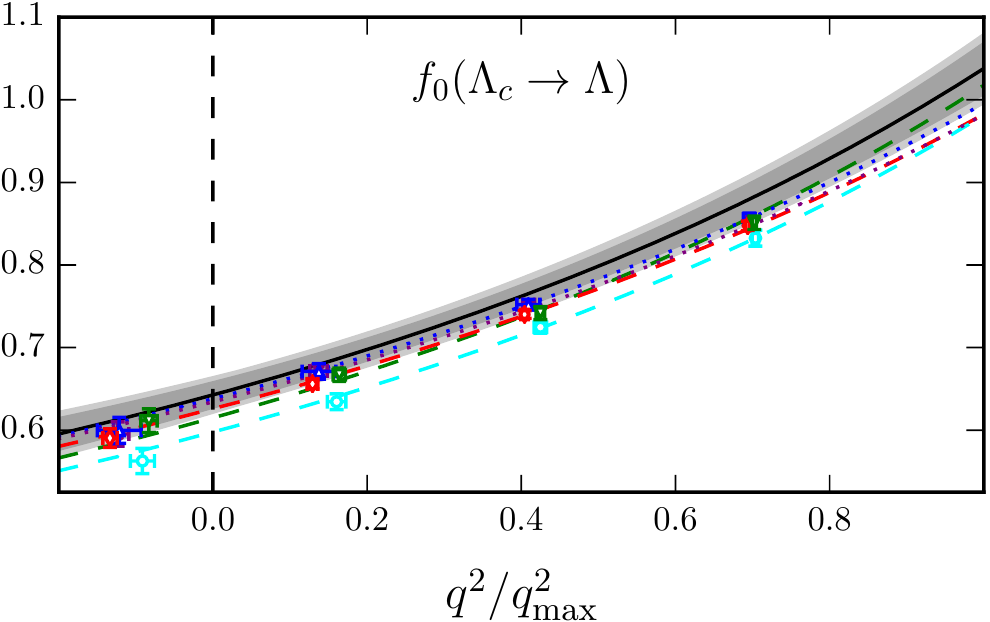} \includegraphics[width=0.45\linewidth]{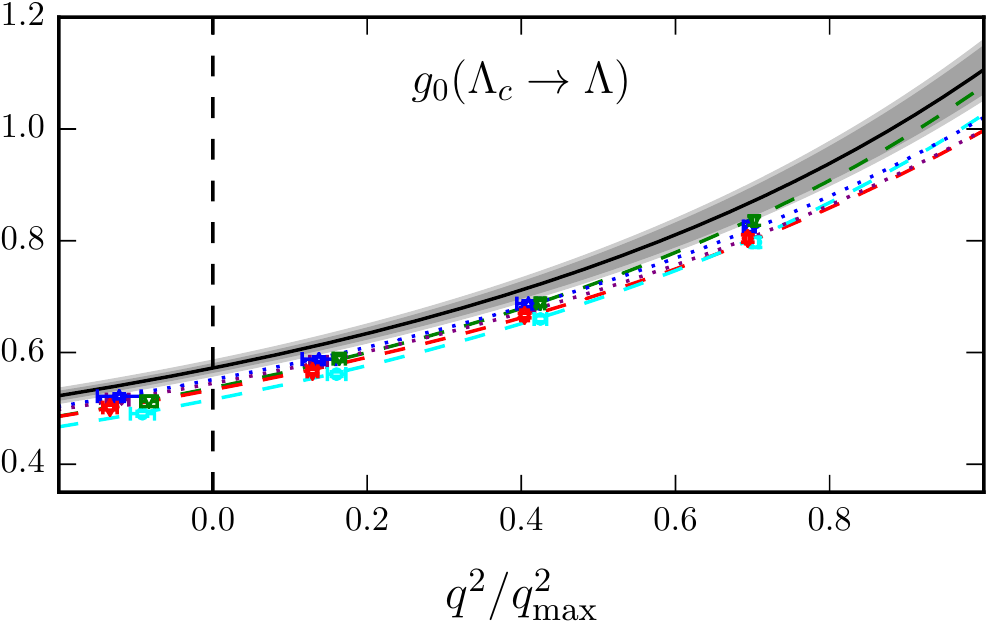} 
\caption{\label{fig:FF}Lattice QCD results for the $\Lambda_c\to\Lambda$ form factors, along with the modified $z$-expansion fits
evaluated at the lattice parameters (dashed and dotted lines) and in the physical limit (solid lines, with statistical and
total uncertainties indicated by the inner and outer bands).} 
\end{figure*}

The ground-state form factors obtained in this way for the different data sets and different discrete momenta are shown as the data points in Fig.~\ref{fig:FF}.
To obtain parametrizations of the form factors in the physical limit ($a=0$, $m_\pi=m_{\pi,{\rm phys}}$, $m_{\eta_s}=m_{\eta_s,{\rm phys}}$),
fits were then performed using $z$-expansions \cite{Bourrely:2008za} modified with additional terms to describe the dependence on $a$, $m_\pi$, and $m_{\eta_s}$.
In the physical limit, the fit functions reduce to the form
\begin{equation}
 f(q^2) = \frac{1}{1-q^2/(m_{\rm pole}^f)^2} \sum_{n=0}^{n_{\rm max}} a_n^f [z(q^2)]^n,
\end{equation}
where $z(q^2)=\displaystyle\frac{\sqrt{t_+-q^2}-\sqrt{t_+-t_0}}{\sqrt{t_+-q^2}+\sqrt{t_+-t_0}}$
with $t_0 = q^2_{\rm max} = (m_{\Lambda_c} - m_{\Lambda})^2$ and $t_+ = (m_D + m_K)^2$. The $D_s$ meson pole masses are $m_{\rm pole}^{f_+,f_\perp}=2.112$ GeV,
$m_{\rm pole}^{f_0}=2.318$ GeV, $m_{\rm pole}^{g_+,g_\perp}=2.460$ GeV,
$m_{\rm pole}^{g_0}=1.968$ GeV \cite{Olive:2016xmw}, and to evaluate $t_+$, the masses $m_D=1.870\:{\rm GeV}$ and $m_K=494\:{\rm MeV}$ are used.
Following Refs.~\cite{Detmold:2015aaa,Detmold:2016pkz}, two separate fits were performed: a ``nominal fit'', giving the central values
and statistical uncertainties of the form factors, and a ``higher-order fit'', used to compute systematic uncertainties according to Eqs.~(50-56) of Ref.~\cite{Detmold:2016pkz}.
The nominal fit had the same form as Eq.~(36) of Ref.~\cite{Detmold:2016pkz}, but with $n_{\rm max}=2$ instead of $n_{\rm max}=1$
(no prior constraints on any parameters were used in the nominal fit). The higher-order fit had the same form as in Eq.~(39) of Ref.~\cite{Detmold:2016pkz},
but with $n_{\rm max}=3$. In addition to the $z^3$ terms, this fit also includes terms of higher order in $a$, $m_\pi$, $m_{\eta_s}$,
and was performed after modifying the data correlation matrix to include the uncertainties from the renormalization and $\mathcal{O}(a)$-improvement
coefficients, from finite-volume effects (1.0\%, rescaled from Ref.~\cite{Detmold:2016pkz} according to $e^{-{\rm min}[m_\pi L]}$), and from the missing isospin breaking/QED corrections (0.5\%, 0.7\%). The priors for the higher-order
parameters were chosen as in Ref.~\cite{Detmold:2016pkz}, except that the coefficients $a_2^f$ were left unconstrained and the priors for $a_3^f$
were set to $0\pm30$. The fit results for the parameters $a_n^f$ that describe the form factors in the physical limit are given in Table \ref{tab:fitresults},
and the form factors are plotted in Fig.~\ref{fig:FF}. The lattice results do not significantly constrain the $z^3$ terms (note that $z_{\rm max}\approx 0.08$),
so that their uncertainties are governed by the priors.

\begin{table}
\begin{tabular}{lllll}
\hline\hline
            & & \hspace{1ex} Nominal fit  &  & Higher-order fit \\
\hline
$a_0^{f_\perp}$     && $\wm 1.30\pm 0.06$ && $\wm 1.28\pm 0.07$ \\ 
$a_1^{f_\perp}$     &&    $-3.27\pm 1.18$ &&    $-2.85\pm 1.34$ \\ 
$a_2^{f_\perp}$     && $\wm 7.16\pm 11.6$ && $\wm 7.14\pm 12.2$ \\ 
$a_3^{f_\perp}$     &&                    &&    $-1.08\pm 30.0$ \\
\hline
$a_0^{f_+}$         && $\wm 0.81\pm 0.03$ && $\wm 0.79\pm 0.04$ \\ 
$a_1^{f_+}$         &&    $-2.89\pm 0.52$ &&    $-2.38\pm 0.61$ \\ 
$a_2^{f_+}$         && $\wm 7.82\pm 4.53$ && $\wm 6.64\pm 6.07$ \\ 
$a_3^{f_+}$         &&                    &&    $-1.08\pm 29.8$ \\ 
\hline
$a_0^{f_0}$         && $\wm 0.77\pm 0.02$ && $\wm 0.76\pm 0.03$ \\ 
$a_1^{f_0}$         &&    $-2.24\pm 0.51$ &&    $-1.77\pm 0.58$ \\ 
$a_2^{f_0}$         && $\wm 5.38\pm 4.80$ && $\wm 4.93\pm 6.28$ \\ 
$a_3^{f_0}$         &&                    &&    $-0.26\pm 29.8$ \\ 
\hline
$a_0^{g_\perp,g_+}$ && $\wm 0.68\pm 0.02$ && $\wm 0.67\pm 0.02$ \\ 
$a_1^{g_\perp}$     &&    $-1.91\pm 0.35$ &&    $-1.73\pm 0.54$ \\ 
$a_2^{g_\perp}$     && $\wm 6.24\pm 4.89$ && $\wm 5.97\pm 6.64$ \\ 
$a_3^{g_\perp}$     &&                    &&    $-1.68\pm 29.8$ \\ 
\hline
$a_1^{g_+}$         &&    $-2.44\pm 0.25$ &&    $-2.22\pm 0.35$ \\ 
$a_2^{g_+}$         && $\wm 13.7\pm 2.15$ && $\wm 12.1\pm 4.43$ \\ 
$a_3^{g_+}$         &&                    && $\wm 12.9\pm 29.2$ \\ 
\hline
$a_0^{g_0}$         && $\wm 0.71\pm 0.03$ && $\wm 0.72\pm 0.04$ \\ 
$a_1^{g_0}$         &&    $-2.86\pm 0.44$ &&    $-2.80\pm 0.53$ \\ 
$a_2^{g_0}$         && $\wm 11.8\pm 2.47$ && $\wm 11.7\pm 4.74$ \\ 
$a_3^{g_0}$         &&                    && $\wm 1.35\pm 29.4$ \\ 
\hline\hline
\end{tabular}
\caption{\label{tab:fitresults}Results for the $z$-expansion parameters describing the form factors in the physical limit. Files containing the parameter values
with more digits and the full covariance matrices are provided as supplemental material \cite{supplementalmaterial}.}
\end{table}

\begin{figure}
\includegraphics[width=\linewidth]{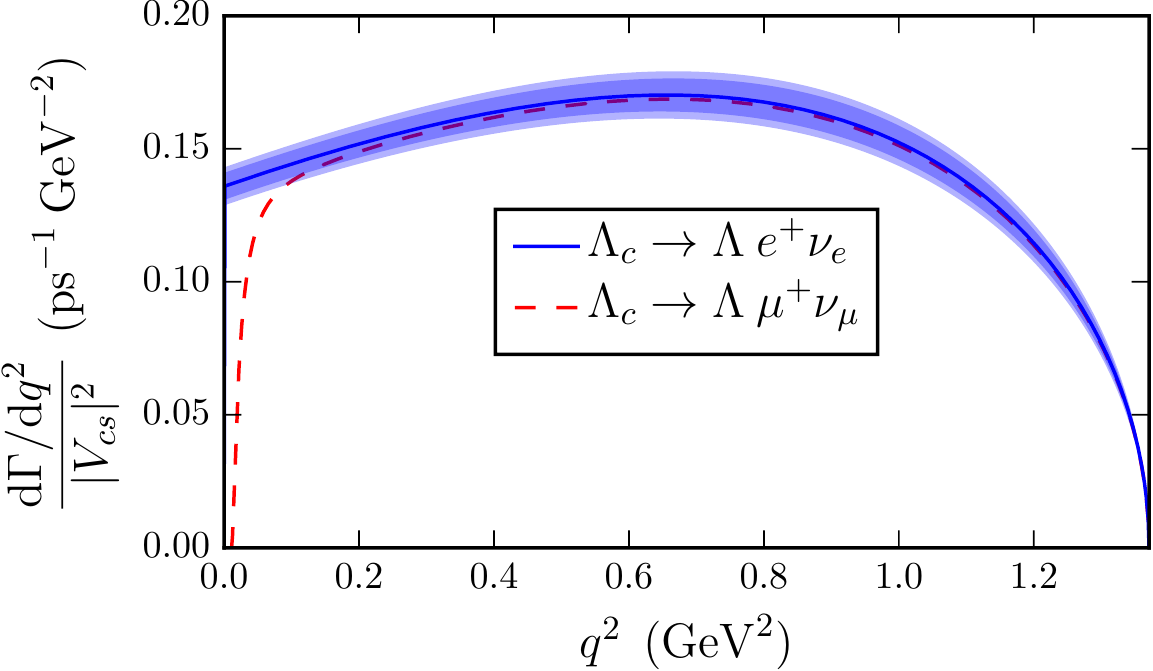} 
\caption{\label{fig:dGamma}Predictions for the $\Lambda_c\to\Lambda\ell^+\nu_{\ell}$ differential decay rates (divided by $|V_{cs}|^2$)
in the Standard Model. For clarity, the uncertainties are shown only for $\ell=e$; the inner and outer bands correspond to
the statistical and total uncertainties. } 
\end{figure}

The resulting Standard-Model predictions for the $\Lambda_c\to\Lambda\ell^+\nu_{\ell}$ differential decay rates,
without the factor of $|V_{cs}|^2$, are shown in Fig.~\ref{fig:dGamma}. The $q^2$-integrated rates are
\begin{equation}
 \frac{\Gamma(\Lambda_c\to\Lambda \ell^+\nu_\ell)}{|V_{cs}|^2}
 = \left\{\begin{array}{ll} 0.2007(71)(74)\:{\rm ps}^{-1}, & \ell=e, \\
                            0.1945(69)(72)\:{\rm ps}^{-1}, & \ell=\mu, \end{array}\right. \label{eq:Gamma}
\end{equation}
where the two uncertainties are from the statistical and total systematic uncertainties in the form factors.
Using the world average of $\Lambda_c$ lifetime measurements, $\tau_{\Lambda_c}=0.200(6)\:{\rm ps}$ \cite{Olive:2016xmw},
and $|V_{cs}|=0.97344(15)$ from a CKM unitarity global fit \cite{UTfit} then yields the branching fractions
\begin{equation}
 \mathcal{B}(\Lambda_c\to\Lambda \ell^+\nu_\ell)
 = \left\{\begin{array}{ll} 0.0380(19)_{\rm LQCD}(11)_{\tau_{\Lambda_c}}, & \ell=e, \\
                            0.0369(19)_{\rm LQCD}(11)_{\tau_{\Lambda_c}}, & \ell=\mu, \end{array}\right.
\end{equation}
where the uncertainties marked ``LQCD'' are the total form factor uncertainties from the lattice calculation. These results are
consistent with, and two times more precise than, the BESIII measurements shown in Eq.~(\ref{eq:BESIII}). This is
a valuable check of the lattice methods which were also used in Refs.~\cite{Aaij:2015bfa, Meinel:2016grj, Detmold:2015aaa, Detmold:2016pkz}.

Combining instead the BESIII measurements (\ref{eq:BESIII}) and $\tau_{\Lambda_c}=0.200(6)\:{\rm ps}$ with the results in Eq.~(\ref{eq:Gamma}) to determine
$|V_{cs}|$ from $\Lambda_c\to\Lambda \ell^+\nu_\ell$ gives
\begin{equation}
 |V_{cs}| = \left\{\begin{array}{ll} 0.951(24)_{\rm LQCD}(14)_{\tau_{\Lambda_c}}(56)_{\mathcal{B}}, & \ell=e, \\
                                     0.947(24)_{\rm LQCD}(14)_{\tau_{\Lambda_c}}(72)_{\mathcal{B}}, & \ell=\mu, \\
                                     0.949(24)_{\rm LQCD}(14)_{\tau_{\Lambda_c}}(49)_{\mathcal{B}}, & \ell=e,\mu, \end{array}\right.
\end{equation}
where the last line is the correlated average over $\ell=e,\mu$. This is the first determination of $|V_{cs}|$ from baryonic decays. The
result is consistent with CKM unitarity, and the uncertainty can be reduced further with more precise measurements of the $\Lambda_c\to\Lambda \ell^+\nu_\ell$
branching fractions.

\begin{acknowledgments}
  \textit{Acknowledgments:} I thank Christoph Lehner for computing the perturbative renormalization and improvement coefficients, and Sergey Syritsyn
  for help with the generation of the domain-wall propagators on the physical-pion-mass ensemble. I am grateful to the RBC and UKQCD Collaborations for making their
  gauge field ensembles available. This work was supported by National Science Foundation Grant No.~PHY-1520996 and by the RHIC Physics Fellow Program
  of the RIKEN BNL Research Center.  High-performance computing resources were provided by the Extreme Science and Engineering Discovery Environment (XSEDE),
  supported by National Science Foundation Grant No.~ACI-1053575, as well as the National Energy Research Scientific Computing Center, a DOE Office of Science
  User Facility supported by the Office of Science of the U.S. Department of Energy under Contract No. DE-AC02-05CH11231.
  The Chroma \cite{Edwards:2004sx} and QLUA \cite{QLUA} software systems were used. Parallel I/O was performed using HDF5 \cite{Kurth:2015mqa}.
\end{acknowledgments}

\providecommand{\href}[2]{#2}
\begingroup
\raggedright

\endgroup

\end{document}